\documentclass{icrc29}
\usepackage{graphicx,amssymb,amsmath,times}
\begin{document}
%Title of paper
\title[The Baikal neutrino...]  {The Baikal neutrino experiment: from NT200 to NT200+ }

\author[Aynutdinov et al ...]  { 
    V. Aynutdinov$^a$, V.Balkanov$^a$, I. Belolaptikov$^g$, N.Budnev$^b$,
    L. Bezrukov$^a$, D. Borschev$^a$,
\newauthor 
    A.Chensky$^b$, I. Danilchenko$^a$, 
    Ya.Davidov$^a$,Zh.-A. Djilkibaev$^a$, G. Domogatsky$^a$, 
\newauthor
A.Dyachok$^b$, 
    S.Fialkovsky$^d$,O.Gaponenko$^a$, O. Gress$^b$, T. Gress$^b$, O.Grishin$^b$, 
    R.Heller$^h$, 
\newauthor 
    A.Klabukov$^a$, A.Klimov$^f$, K.Konischev$^g$, A.Koshechkin$^a$, 
     L.Kuzmichev$^c$, V.Kulepov$^d$, 
\newauthor  
    B.Lubsandorzhiev$^a$, S.Mikheyev$^a$, 
    M.Milenin$^d$, R.Mirgazov$^b$, T.Mikolajski$^h$, E.Osipova$^c$,
\newauthor 
    A.Pavlov$^b$,  G.Pan'kov$^b$, L.Pan'kov$^b$, A.Panfilov$^a$,
    \framebox{Yu.Parfenov$^b$}, D.Petuhov$^a$,
\newauthor 
    E.Pliskovsky$^g$, P.Pokhil$^a$, V.Polecshuk$^a$, E.Popova$^c$, V.Prosin$^c$, 
    M.Rozanov$^e$, 
\newauthor 
    V.Rubtzov$^b$, B.Shaibonov$^a$, A.Shirokov$^c$,Ch. Spiering$^h$, 
    B.Tarashansky$^b$,
 \newauthor 
    R.Vasiliev$^g$, E.Vyatchin$^a$, R.Wischnewski$^h$,
    I.Yashin$^c$, V.Zhukov$^a$ \\ 
 (a) Institute for Nuclear Reseach, Russia\\
 (b) Irkutsk State University,Russia\\
 (c) Skobeltsin Institute of Nuclear Physics, Moscow State University, Russia\\
 (d) Nizni Novgorod State Technical University, Russia\\
 (e) St.Petersburg State Marine Technical University, Russia\\
 (f) Kurchatov Institute, Russia\\
 (g) Joint Institute for Nuclear Research, Dubna, Russia\\
 (h) DESY, Zeuthen, Germany}

\presenter{Presenter: Kuzmichev L.A. (kuz@dec1.sinp.msu.ru), \  
rus-kuzmichev-lA-abs3-og25-oral}

\maketitle

\begin{abstract}
 The Baikal neutrino telescope NT200 takes data since 1998. In 2005, the deployment of three
 additional strings for common operation with NT200 was finished. We describe the physics
 program  and the design of the new telescope named NT200+ and present selected physical
 results obtained with NT200. First results from NT200+ will be presented at the conference.

\end{abstract}
\protect\vspace*{-2mm}
\section{Introduction}
The Baikal Neutrino Telescope is operated in Lake Baikal, Siberia, at a depth of 1.1 km,
3.6 km from the shore.
Deep Baikal water is characterized by an absorption length of $L_{abs}(480 $nm$) =20\div 24$ m,
a scattering length of $L_s -30\div 70$ m and a strongly anisotropic scattering function
with a mean cosine of scattering angle $0.85\div 0.9$. 
The present stage of telescope, NT200+ (fig.\ref{nt_pl}) was put into operation at April 9th, 2005.
The new configuration consists of the old NT200 telescope plus the three new, external strings, 
placed 100 m from
the center of NT200. With the new strings, the sensitivity of Baikal telescope 
for very high energy neutrinos increases by a factor 4.  
Below we describe the new telescope and present selected physical and methodical results.
Results on the search for high energy neutrinos and relativistic monopoles are presented in
separate papers.

\protect\vspace*{-2mm}
\section{NT200+}

  NT200 plus the new external strings form NT200+.
  NT200 \cite{NT200} with 192 optical modules(OM), was put into operation in 1998. 
  An umbrella-like frame carries 8 strings 72 m long, each with 24 pairwise arranged OMs.
  The optical modules contain 37-cm photodetectors QUASAR-370 \cite{QUASAR}, which have been developed
  specially for our project. The OMs are grouped in pairs along the strings. The PMs of a pair are
  switched in coincidence in order to suppress background from bioluminescence and PM noise. 
  A pair defines a \textit{channel}.

\begin{figure}[t]
\begin{center}
\includegraphics*[width=0.42\textwidth, trim =3mm 0mm 0mm 15 mm,angle=0,clip]{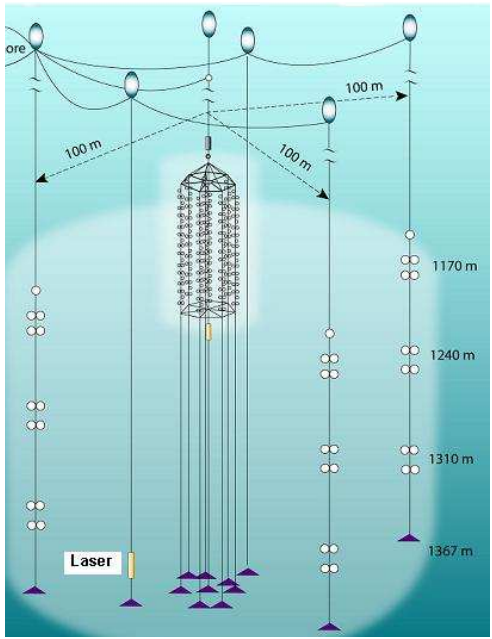}
\vspace*{-3mm}
\caption{\label{nt_pl} Sketch of NT200+}
\end{center}
\vspace*{-10 mm}
\end{figure}

\vspace*{-3mm}
A \textit{trigger} is formed by requiring $\geq N$\textit{hits}( with \textit{hit} refering to
a channel) within 500 ns. $N$ is typically set to 3 or 4. This signals is used as a \textit{common}
stop for TDCs.
For the common operation with the external
strings the signal of the \textit{trigger} is sent through 1.2 km coaxial cable  
to the {\it underwater} DAQ center of 
NT200+ (20 m below surface, distributed over several glass spheres). 
For time calibration of NT200, two nitrogen lasers, placed on the top and below the central 
string, are used. The light pulse from the first laser ($\lambda =$ 480 nm) is sent 
to all OMs via optical fibers with equal length, the
light pulse from the second is emitted  straight to the water.

External strings 140 m long are placed at 100 meter distance 
from the center of NT200, each with 12 OMs grouped in
pairs like in NT200. The upper pairs are at approximately the same level as the bottom OMs of NT200. 
On each external string, an  independent  \textit{string trigger} is formed in case the
number of fired channels is
$\geq$2 within 1000 ns. \textit{String triggers} are sent to the underwater DAQ center, where the time
difference between \textit{string trigger} and the \textit{trigger} of NT-200 is measured. This
information is used to relate the times within an event of NT200 OMs and OMs of the externals strings.
For time offset synchronization of the external strings with respect to
NT200, an additional laser, placed
on a separate string (see fig.\ref{nt_pl} left bottom) is used. 
This laser emits ten times more photons ($\ge 10^{12}$)  
than the old ones and fires OMs on all externals strings as well as on NT200. 
With this laser it was found that the jitter of the time difference
between OMs of NT200 and OMs of the external strings is smaller than 3 ns.
  
\begin{figure}[h]
\includegraphics[width=0.35\textwidth, angle=0,clip]{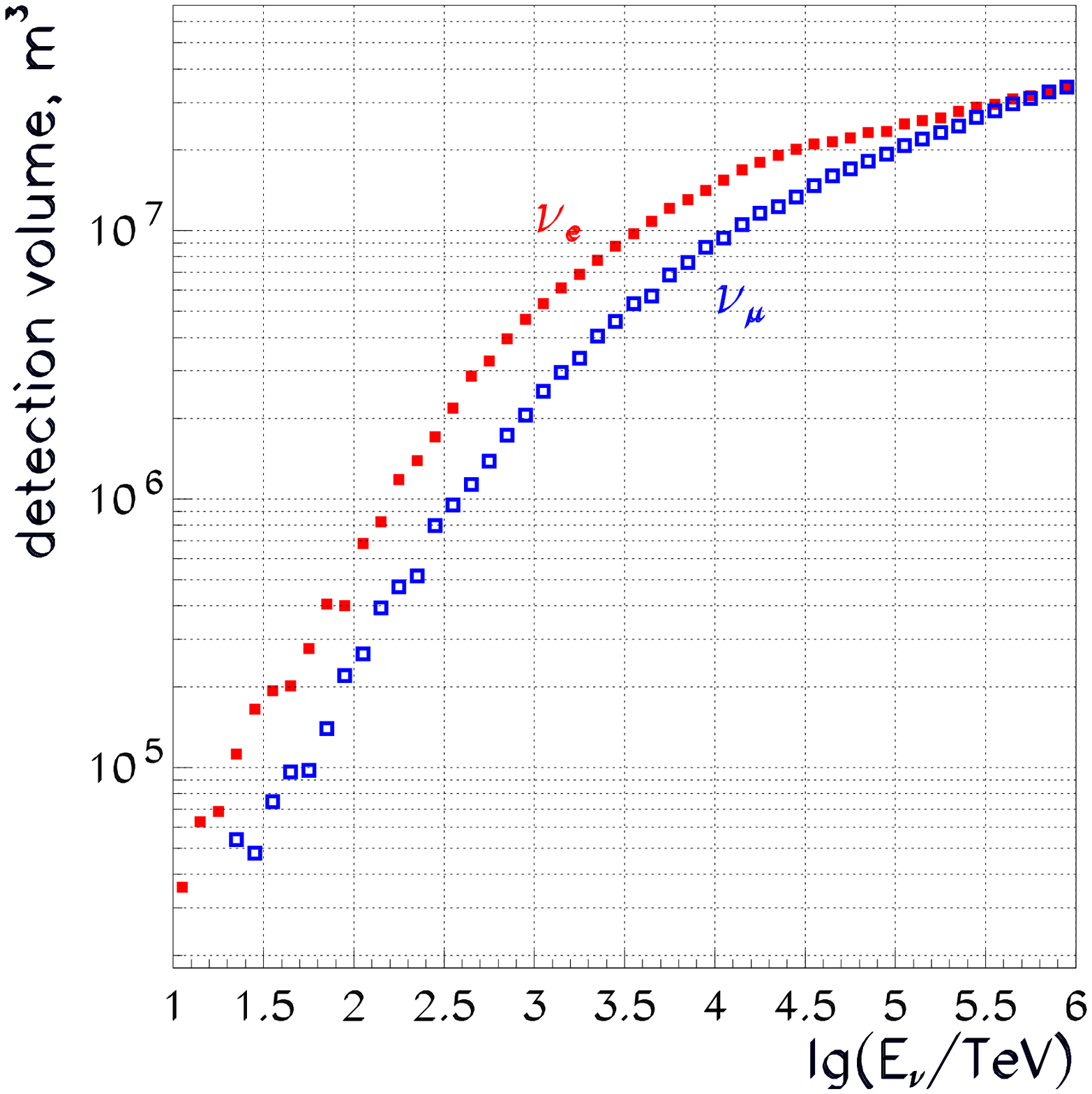}
\hfil
\includegraphics[width=0.35\textwidth, angle=0,clip]{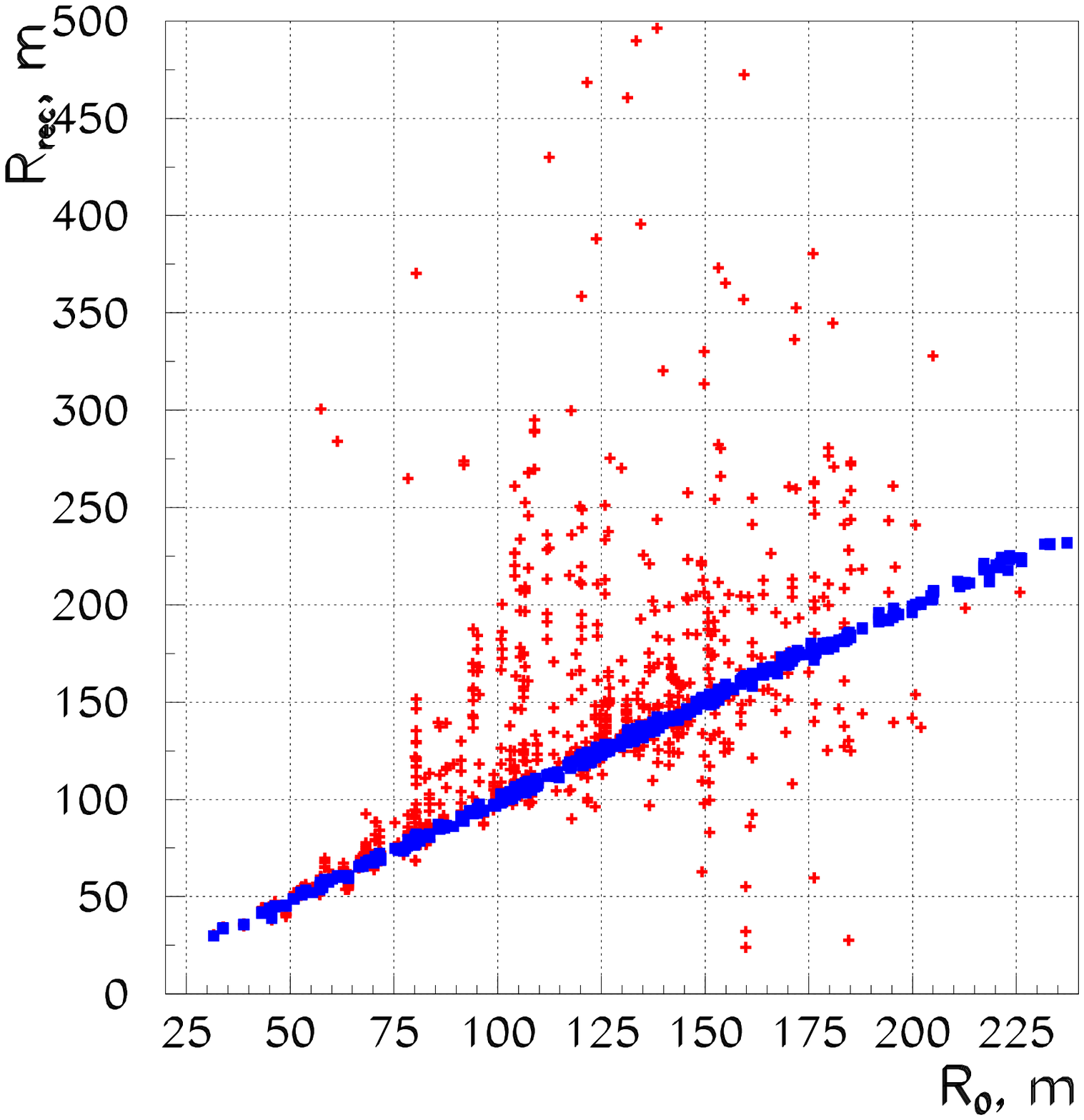}
\\
\parbox[h]{0.47\textwidth}{\caption{Detection volume of NT200+ for $\nu_e$ and
$\nu_{\mu}$ events which survives all cuts}\label{v131}}
\hfil
\parbox[h]{0.47\textwidth}{\caption{ Reconstructed v. simulated coordinates of cascades in
NT-200+ (rectangles) and NT-200(crosses)}\label{map}}
\end{figure} 

Our search strategy for high energy neutrinos relies on the detection of the Cherenkov light emitted
by cascades, produced by neutrino interactions in a large volume mostly below NT200.
The detection volume for NT200+ is shown in fig.\ref{v131}. With three additional strings,
it is increased by four times. With NT200+ we also significantly increase 
the accuracy of the position of the cascade vertex
 (fig.\ref{map}) within the volume spanned by the outer strings, and based on
that the energy reconstruction.  

For the first 1000 hours of operation, $7.5\cdot10^4$ common events (detected by NT200 and
external strings) have been collected. Data analysis is in progress and first results will be
presented at the conference.

\protect \vspace*{-5 mm}
\section{Selected results}  
\vspace*{-1 mm}
\subsection{Atmospheric Muon Neutrinos}
\vspace*{-2 mm}
The signature of charged current muon neutrino events is a muon crossing the detector
from below. Muon track reconstruction algorithms and background rejection have 
been described elsewhere \cite{rec-nu}. Compared to \cite{rec-nu},
the analysis of the 5-year sample
(1038 days for pure operation time) 
was optimized for higher signal passing rate, and accepting a slightly higher
contamination of 15-20\% fake events. A total of 372 upgoing neutrino 
candidates were found. From
Monte-Carlo simulations, a total 385 atmospheric neutrino and 
background events are expected. The skyplot of these events is shown in fig.\ref{fig3}.

\vspace*{-2mm}
\subsection{Search for Neutrinos from WIMP Annihilation}
\begin{figure}[t]
\includegraphics[width=0.47\textwidth,angle=0,clip]{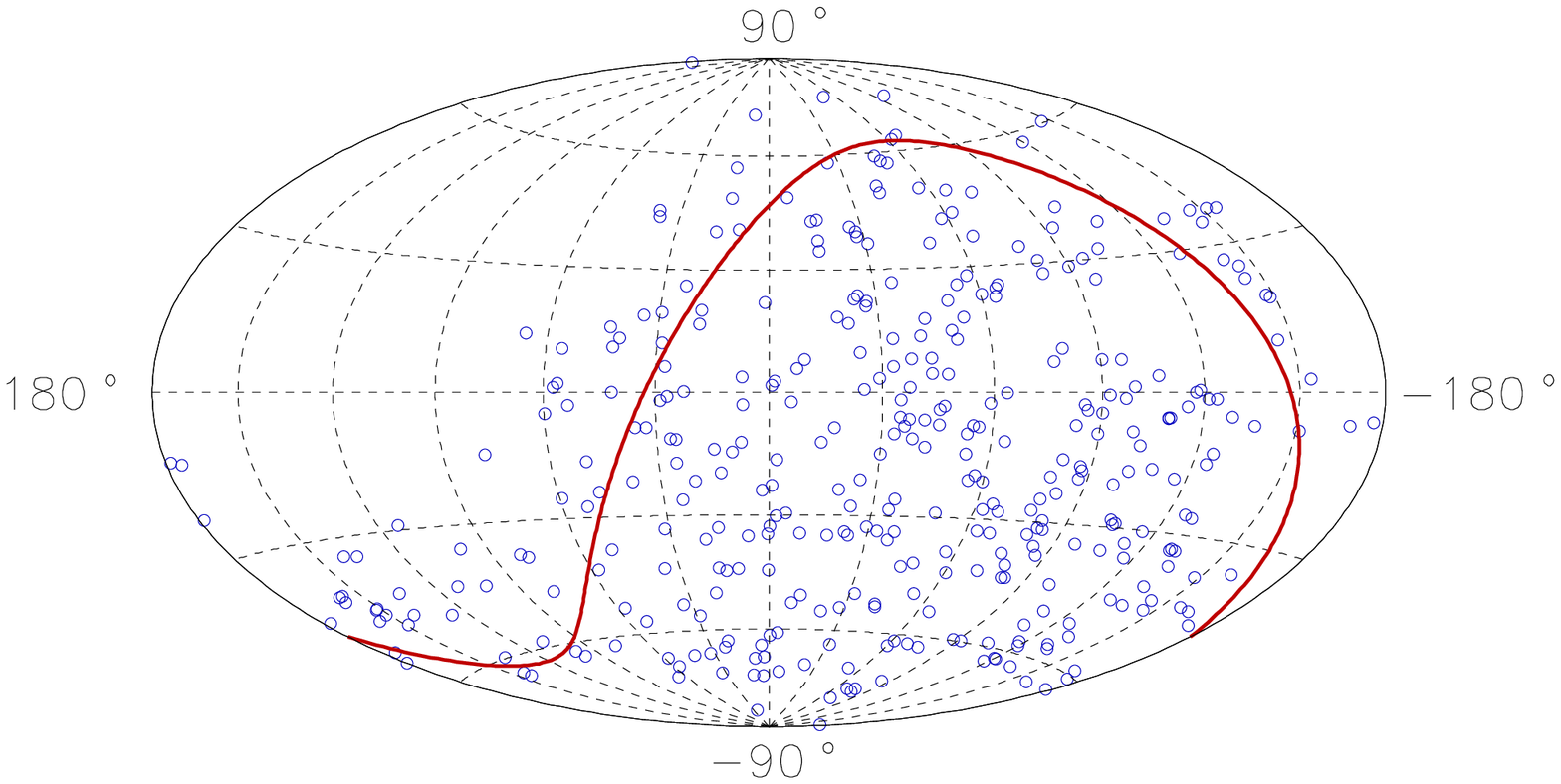}
\hfil
\includegraphics[width=0.47\textwidth, angle=0,clip]{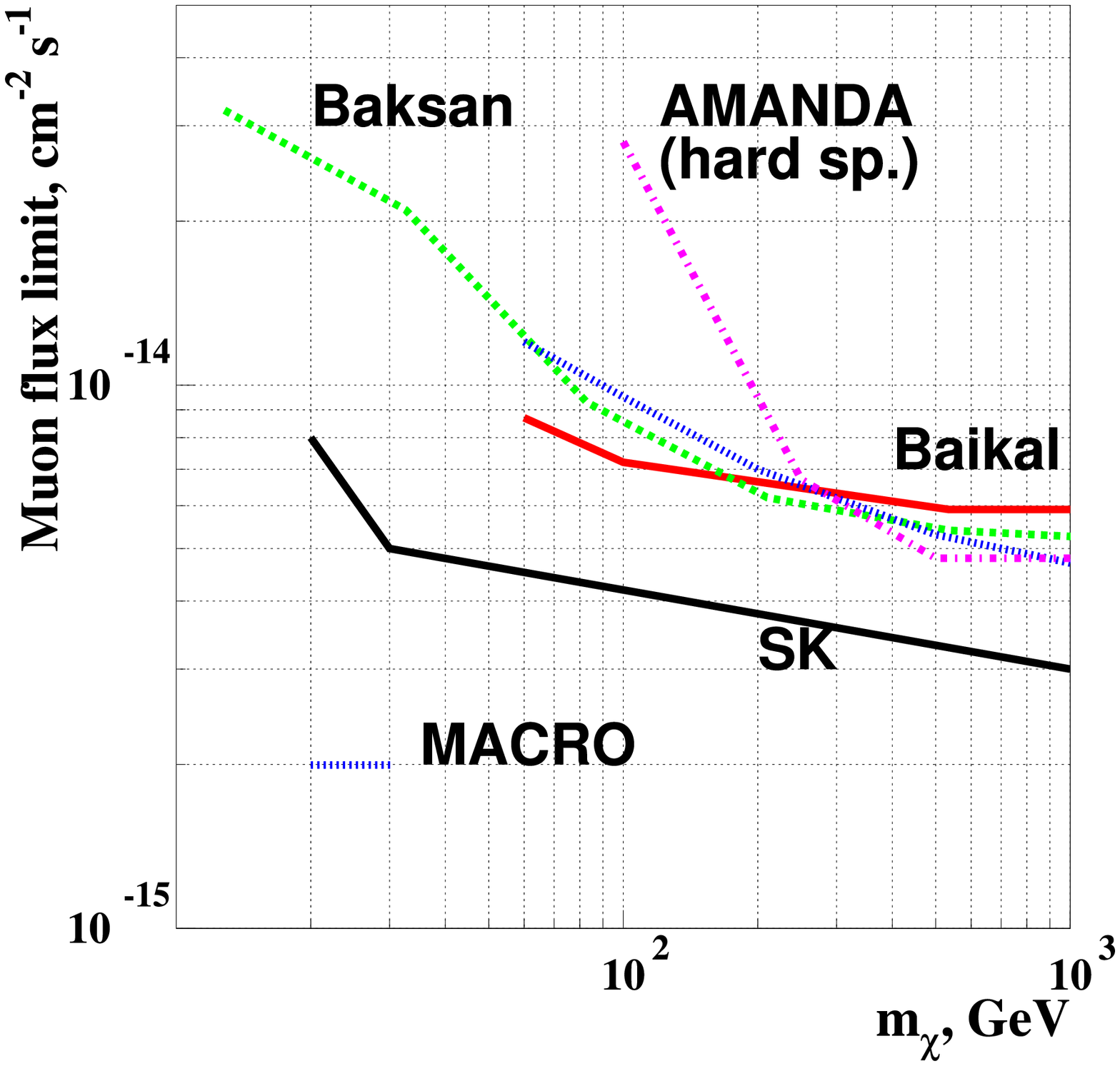}
\vspace*{-3mm}
\parbox[t]{0.47\textwidth}
{\caption{Skyplot of neutrino events in galactic coordinates
for five years}\label{fig3}}
\hfil
%\vspace*{-5mm}
\parbox[t]{0.47\textwidth}
{\caption{Limits on the excess muon flux from the
center of the Earth as function of the WIMP mass, Reference to other limits, presented on this figure,
maybe found in $[$6$]$}\label{fig4}}
\end{figure}

The search for WIMPs with the Baikal neutrino telescope is based on a 
possible signal of nearly vertically upward going muons, exceeding the flux of
atmospheric neutrinos. The applied cuts for event selection \cite{wimp} 
result in a detection area of 1800 m$^2$ for vertically upward going muons.
In 502 days
effective data taking, 24 events with $-0.75 \ge \cos(\theta) \ge -1$ have been found,
in accordance with expectations for atmospheric neutrinos.
Regarding these 24 events as being induced by atmospheric neutrinos, one can derive
an upper limit on the excess flux of muons from the center of the Earth due to
annihilation of neutralinos (fig.\ref{fig4}).

\vspace*{-2mm}
\subsection{ Search for acoustic signals at Lake Baikal}            
 In 2001--2003, a series of experiments has been performed to detect acoustic signals 
 in water from
 EAS. For these experiments, a small EAS array was deployed on the ice. 
 A special acoustic device recorded sound  
 signals for a 100 ms interval after the EAS time.
 This device consists of 4 hydrophones arranged at the vertices of a pyramide 
  with 1.5 m side length.
 During these experiments many bipolar signals from close to the EAS core
  were detected, but due the very high level of acoustic noise we cannot prove that
  these signals are really from EAS\cite{GVD}. In 2005,  four hydrophones  with
  a new data acquisition system were tested in situ. With this device it is
  possible to reconstruct the direction to the sound source with
  one degree accuracy. During a 20-hour test with the device at
  100 m depth it was found that the majority of the acoustic signals comes from
  the surface of the lake and only a few from below the horizon.
  No signals from directions steeper than 20$^0$ below horizon
  have been detected. In 2006 we plan to operate
  this device together with NT200+.
     
\vspace*{-3mm} 
\section{Conclusion and Outlook}
\vspace*{-1mm}
The Baikal Neutrino telescope NT200, operated from 1998,  
 was upgraded to NT200+, a 10-Mton detector with a
 sensitivity of approximately $\Phi_{\nu_e}E^2\sim$ 10$^{-7}$ cm$^{-2}$ sr$^{-1}$GeV
 for diffuse neutrino fluxes at $E \ge 10^2$ TeV.
 NT200+ will search for neutrinos from AGNs, GRBs and other extraterrestrial
 sources, neutrinos from cosmic ray interactions in the Galaxy as well as high energy
 atmospheric muons with $E_{\mu} \ge 10$ TeV,and also for exotic particles (monopoles, strangles, Q-balls).
  In parallel to this short term goal,
 we started research $\&$ development activities towards a Gigaton Volume 
 Detector \cite{GVD} in Lake Baikal. Our last results on the search for high energy
 neutrinos and relativistic monopoles and more detailed descriptions of the data acquisition
 system of NT200+ are presented in  separate papers on this conference \cite{rf1, rf2}.\\
 \textit{This work was supported by the Russian Ministry of Education and Science, the 
 German Ministry of Education and Research and the Russian Fund of Basic Research
 (gratns 05-02-17476, 04-02-17289, 04-02-16171, 05-02-16593), by the Grant of
 the President of Russia NSh-1828.2003.2, and by NATO-Grant NIG-981707}


\vspace*{-2mm}
\begin{thebibliography}{99}

\bibitem{NT200}
I.~Belolaptikov et al., Astropart.Phys. 7, 263 (1997).


\bibitem{QUASAR}
R.~I.~Bagduev et al., Nucl.Instr.Meth. A420, 138(1999).

\bibitem{rec-nu}
I.~Belolaptikov et al., Astropart.Ph. 12, 75(1999)

\bibitem{GVD}
V.~Balkanov et al.,  Nucl. Phys.B (Proc.Suppl.)118, 363(2003)

\bibitem{wimp}
V.~Balkanov et al., Nucl.Phys.(Proc.Suppl.) 91, 438(2000) 

\bibitem{cs}
 C.~Spiering.  astro-ph/0503122 (2005)

\bibitem{rf1}
V.~Aynutdinov et al.[Baikal Collaboration], these proceedings: ger-wischnewski-R-abs1-og25-oral

\bibitem{rf2}
V.~Aynutdinov et al.[Baikal Collaboration], these proceedings: ger-wischnewski-R-abs1-he23-poster

\end{thebibliography}
\end{document}